\documentclass[prd,twocolumn,superscriptaddress,showpacs,amsmath,amssymb]{revtex4}
\usepackage{graphicx}
\begin{document}
\title{Cosmological model with viscosity media (dark fluid) described by an
effective equation of state}
\author{Jie Ren}
\email{jrenphysics@hotmail.com} \affiliation{Department of physics,
Nankai University, Tianjin 300071, China (post address)}
\author{Xin-He Meng}
\altaffiliation{Communication author} \email{xhm@nankai.edu.cn}
\affiliation{Department of physics, Nankai University, Tianjin
300071, China (post address)} \affiliation{CCAST (World Lab),
P.O.Box 8730, Beijing 100080, China}
\date{\today}
\begin{abstract}
A generally parameterized equation of state (EOS) is investigated
in the cosmological evolution with bulk viscosity media modelled
as dark fluid, which can be regarded as a unification of dark
energy and dark matter. Compared with the case of the perfect
fluid, this EOS has possessed four additional parameters, which
can be interpreted as the case of the non-perfect fluid with
time-dependent viscosity or the model with variable cosmological
constant. From this general EOS, a completely integrable dynamical
equation to the scale factor is obtained with its solution
explicitly given out. (i) In this parameterized model of
cosmology, for a special choice of the parameters we can explain
the late-time accelerating expansion universe in a new view. The
early inflation, the median (relatively late time) deceleration,
and the recently cosmic acceleration may be unified in a single
equation. (ii) A generalized relation of the Hubble parameter
scaling with the redshift is obtained for some cosmology
interests. (iii) By using the SNe Ia data to fit the effective
viscosity model we show that the case of matter described by $p=0$
plus with effective viscosity contributions can fit the
observational gold data in an acceptable level.
\end{abstract}
\maketitle

\section{Introduction}
The cosmological observations have provided increasing evidence
that our universe is undergoing a late-time cosmic acceleration
expansion \cite{bah99}. In order to explain the acceleration
expansion, cosmologists introduce a new fluid, which possesses a
negative enough pressure, called dark energy. According to the
observational evidence, especially from the Type Ia Supernovae
\cite{rie04} and WMAP satellite missions\cite{ben03}, we live in a
favored spatially flat universe consisting approximately of $30\%$
dark matter and $70\%$ dark energy. The simplest candidate for
dark energy is the cosmological constant, but it has got the
serious fine-tuning problem. Recently, a great variety of models
are proposed to describe the universe with dark energy, partly
such as
\begin{itemize}
\item Scalar fields: Quintessence \cite{wan99} and phantom
\cite{cal99}, the model potential is from power-law to
exponentials and a combination of both. \item Exotic equation of
state: Chaplygin gas \cite{kam01}, generalized Chaplygin gas
\cite{ben02}, a linear equation of state \cite{bab05}, and Van der
Waals equation of state \cite{cap02}. \item Modified gravity: DGP
model \cite{dgp}, Cardassian expansion \cite{fre02}, $1/R$, $R^2$,
ln$R$ term corrections, etc. Maybe the mysterious dark energy does
not exist, but we lack the full understanding of gravitational
physics \cite{chi03,fla04,noj03,vol03,lue03,car04,noj04a,mw1}.
\item Viscosity: Bulk viscosity in the isotropic space \cite{tp},
bulk and shear viscosity in the anisotropic space. The perfect
fluid is only an approximation of the universe media. The
observations also indicate that the universe media is not a
perfect fluid \cite{jaf05} and the viscosity is concerned in the
evolution of the universe \cite{bre05a,bre05b,cat05}.
\end{itemize}
We only list a part of the papers on this topics as the relevant
are too many. According to Ref.~\cite{cap05},it is possible to put
some order in this somewhat chaotic situation by considering a
particular feature of the dark energy, namely its equation of
state (hereafter EOS); it is tempting to investigate the
properties of cosmological models starting from the EOS directly
and by testing whether a given EOS is able to give rise to
cosmological models reproducing the available dataset. The dark
fluid \cite{arb05} and the parameterized EOS \cite{joh05} are
studied in some recent papers. We hope the situation will be
improved with the new generation of more precise observational
data.

The observational constraints indicate that the current EOS
parameter $w=p/\rho$ is around $-1$ \cite{rie04, tps}, quite probably
below $-1$, which is called the phantom region and even more
mysterious in the cosmological evolution. In the standard model of
cosmology, if the $w<-1$, the universe shows to possess the future
finite singularity called Big Rip \cite{cal03,noj05a}. Several
ideas are proposed to prevent the big rip singularity, like by
introducing quantum effects terms in the action \cite{noj04b}.

Based on the motivations of time-dependent viscosity and modified
gravity, the Hubble parameter dependent EOS is considered in
Ref.~\cite{noj05b,cap05}, in which the most general
"inhomogeneous" EOS is given out, however, they gives analytical
solutions of the scale factor only for some less general cases. In
this paper, we investigate a general effective equation of state
$$p=(\gamma-1)\rho+p_0+w_{H}H+w_{H2}H^2+w_{dH}\dot{H},$$
and we show the following time-dependant bulk viscosity
$$\zeta=\zeta_0+\zeta_1\frac{\dot{a}}{a}+\zeta_2\frac{\ddot{a}}{\dot{a}}$$
is equivalent to the form derived by using the above effective
EOS. An integrable equation for the scale factor is obtained and
three possible interpretations of this equation are proposed. The
Hubble parameter dependent term in this EOS can drive the phantom
barrier being crossed in an easier way \cite{bre05a,noj05b,men05}.
Different choices of the parameters may lead to several  fates to
the cosmological evolution \cite{men05}.

This paper is organized as follows: In the next section we
describe our model and give the exact solution of the scale
factor. In Sec. III we consider the sound speed and the EOS
parameter in this model for unified dark energy, and give some
numerical solutions of a more general equation for the scale
factor. In Sec. IV we propose three interpretations for our model.
In Sec. V  we confront the effective viscosity model proposed in
the previous Sec. with the SNe Ia Golden data. Finally, we present
our conclusions in the last section. The appendix presents some detail
comments on the cosmological constant involved.

\section{Model and calculations}
We consider the Friedamnn-Roberson-Walker metric in the flat space
geometry ($k=0$) as favored by WMAP cosmic microwave  background
data on power spectrum
\begin{equation}
ds^2=-dt^2+a(t)^2(dr^2+r^2d\Omega^2),
\end{equation}
and assume that the cosmic fluid possesses a bulk viscosity
$\zeta$. The energy-momentum tensor can be written as
\begin{equation}
T_{\mu\nu}=\rho U_\mu U_\nu+(p-\zeta\theta)H_{\mu\nu},
\end{equation}
where in comoving coordinates $U^\mu=(1,0)$,
$\theta=U^\mu_{;\mu}=3\dot{a}/a$, and $H_{\mu\nu}=g_{\mu\nu}+U_\mu
U_\nu$ \cite{bre02}. By defining the effective pressure as
$\tilde{p}=p-\zeta\theta$ and from the Einstein equation
$R_{\mu\nu}-\frac{1}{2}g_{\mu\nu}R=\kappa^2 T_{\mu\nu}$ with
$\kappa^2=8\pi G$, we obtain the Friedmann equations
\begin{subequations}
\begin{eqnarray}
\frac{\dot{a}^2}{a^2} &=& \frac{\kappa^2}{3}\rho\label{eq:f1},\\
\frac{\ddot{a}}{a} &=&
-\frac{\kappa^2}{6}(\rho+3\tilde{p})\label{eq:f2}.
\end{eqnarray}
\end{subequations}
The conservation equation for energy, $T^{0\nu}_{;\nu}$, yields
\begin{equation}
\dot{\rho}+(\rho+\tilde{p})\theta=0.
\end{equation}

To describe completely the global behaviors for our Universe
evolution an additional relation, a reasonable EOS is required. A
generally parameterized EOS can be written as
\begin{equation}
p=(\gamma-1)\rho+f(\rho;\alpha_i)+g(H,\dot{H};\alpha_i)
\end{equation}
where $\alpha_i$ are parameters that are expected that when
$\alpha_i\to 0$, the equation of state approaches to that of the
perfect fluid, i.e. $p=w\rho$, where the factorized parameter
$w=\gamma-1$ with $\gamma$ being another parameter. We consider
the following EOS, an explicit form as
\begin{equation}
p=(\gamma-1)\rho+p_0+w_{H}H+w_{H2}H^2+w_{dH}\dot{H},\label{eq:eos}
\end{equation}
where $p_0$, $w_{H}$, $w_{H2}$, $w_{dH}$ are free parameters. In
this and the next section, we assume the universe media is a
single fluid described by this EOS. Compared with the bulk
viscosity form as described in Ref.~\cite{men05}, the following
one is more general. We show that this time-dependent bulk
viscosity
\begin{equation}
\zeta=\zeta_0+\zeta_1\frac{\dot{a}}{a}+\zeta_2\frac{\ddot{a}}{\dot{a}}
\end{equation}
is effectively equivalent to the form derived by using
Eq.~(\ref{eq:eos}). The reason is
\begin{eqnarray}
\tilde{p} &=& p-\zeta\theta\nonumber\\
&=& p-3\zeta_0\frac{\dot{a}}{a}-3\zeta_1\frac{\dot{a}^2}{a^2}
-3\zeta_2\frac{\ddot{a}}{a}\nonumber\\
&=&
p-3\zeta_0\frac{\dot{a}}{a}-3(\zeta_1+\zeta_2)\frac{\dot{a}^2}{a^2}
-3\zeta_2\left(\frac{\ddot{a}}{a}-\frac{\dot{a}^2}{a^2}\right)\nonumber\\
&=& p-3\zeta_0 H-3(\zeta_1+\zeta_2)H^2-3\zeta_2\dot{H},
\end{eqnarray}
we can see that the corresponding coefficients are
\begin{subequations}
\begin{eqnarray}
w_{H} &=& -3\zeta_0,\\
w_{H2} &=& -3(\zeta_1+\zeta_2),\\
w_{dH} &=& -3\zeta_2.
\end{eqnarray}
\end{subequations}
The motivation of considering this bulk viscosity is that by fluid
mechanics we know the transport/viscosity phenomenon is related to
the "velocity" $\dot{a}$, which is related to the Hubble
parameter, and the acceleration. Since we do not know the exact
form of viscosity, here we consider a parameterized bulk
viscosity, which is a linear combination of three terms: the first
term is a constant $\zeta_0$, the second corresponds to the Hubble
parameter, and the third can be proportional to $\ddot{a}/aH$.
From the above corresponding coefficients, we can see that the
inhomogeneous EOS may be interpreted simply as time-dependent
viscosity case. Additionally, the EOS of Eq.~(\ref{eq:eos}) can
also be interpreted as the case of a variable cosmological
constant model, in which the $\Lambda$-term is written as
\begin{equation}
\Lambda=\Lambda_0+\Lambda_H H+\Lambda_{H2}H^2+\Lambda_{dH}\dot{H}.
\end{equation}

\begin{figure}[]
\includegraphics{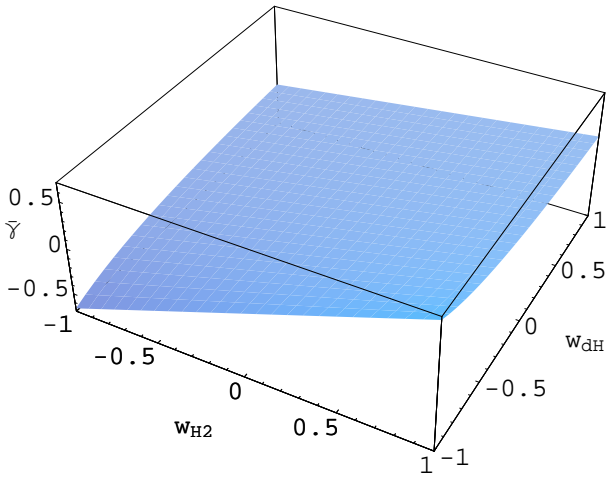}
\caption{\label{gamma1} The relation of $\tilde{\gamma}$, $w_{H2}$,
and $w_{dH}$ when $\gamma=0$.}
\end{figure}

\begin{figure}[]
\includegraphics{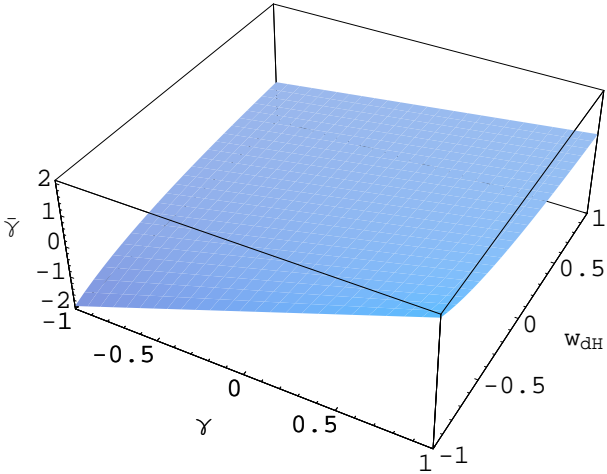}
\caption{\label{gamma2} The relation of $\tilde{\gamma}$, $\gamma$,
and $w_{H2}$ when $w_{H2}=0$}
\end{figure}

Using this EOS to eliminate $\rho$ and $p$, we obtain the equation
which determines the scale factor $a(t)$ evolution
\begin{widetext}
\begin{equation}
\frac{\ddot{a}}{a}=\frac{-(3\gamma-2)/2-(\kappa^2/2)w_{H2}+(\kappa^2/2)w_{dH}}
{1+(\kappa^2/2)w_{dH}}\frac{\dot{a}^2}{a^2}+\frac{-(\kappa^2/2)
w_H}{1+(\kappa^2/2)w_{dH}}\frac{\dot{a}}{a}
+\frac{-(\kappa^2/2)p_0}{1+(\kappa^2/2)w_{dH}}.\label{eq:orig}
\end{equation}
To make this equation more comparable to that of the perfect fluid,
we define $\tilde{\gamma}$ given by
\begin{equation}
\frac{-(3\gamma-2)/2-(\kappa^2/2)w_{H2}
+(\kappa^2/2)w_{dH}}{1+(\kappa^2/2)w_{dH}}=-\frac{3\tilde{\gamma}-2}{2}.
\end{equation}
This equation gives
\begin{equation}
\tilde{\gamma}=\frac{\gamma+(\kappa^2/3)w_{H2}}{1+(\kappa^2/2)w_{dH}}.
\end{equation}
By defining
\begin{eqnarray}
\frac{1}{T_1} &=& \frac{-(\kappa^2/2)
w_H}{1+(\kappa^2/2)w_{dH}}\\
\frac{1}{T_2^2} &=& \frac{-(\kappa^2/2)p_0}{1+(\kappa^2/2)w_{dH}},\\
\frac{1}{T^2} &=& \frac{1}{T_1^2}+\frac{6\tilde{\gamma}}{T_2^2}.
\end{eqnarray}
and noting that dim[$T_1$]=dim[$T_2$]=[time], we can see that when
$T_2\rightarrow\infty$, $T=T_1$; when $T_1\rightarrow\infty$,
$T=T_2\sqrt{6\tilde{\gamma}}$. Now Eq.~(\ref{eq:orig}) becomes
\begin{equation}
\frac{\ddot{a}}{a}=-\frac{3\tilde{\gamma}-2}{2}\frac{\dot{a}^2}{a^2}
+\frac{1}{T_1}\frac{\dot{a}}{a}+\frac{1}{T_2^2}.\label{eq:main}
\end{equation}
The five parameters $\gamma$, $p_0$, $w_H$, $w_{H2}$, and $w_{dH}$
are condensed to three parameters $\tilde{\gamma}$, $T_1$, and
$T_2$ in the above equation.

With the initial conditions of $a(t_0)=a_0$ and
$\theta(t_0)=\theta_0$, if $\tilde{\gamma}\neq 0$, the solution can
be obtained as
\begin{equation}
a(t) = a_0\left\{\frac{1}{2}\left(1+\tilde{\gamma}\theta_0
T-\frac{T}{T_1}\right){\rm{exp}}\left[\frac{t-t_0}{2}\left(\frac{1}{T}
+\frac{1}{T_1}\right)\right]+\frac{1}{2}\left(1-\tilde{\gamma}\theta_0
T+\frac{T}{T_1}\right){\rm{exp}}\left[-\frac{t-t_0}{2}\left(\frac{1}{T}
-\frac{1}{T_1}\right)\right]\right\}^{2/3\tilde{\gamma}},\label{eq:a}
\end{equation}
And we obtain directly
\begin{equation}
\rho(t)=\frac{3}{\kappa^2}\frac{\dot{a}^2}{a^2}=\frac{1}{3\kappa^2\tilde{\gamma}^2}
\left[\frac{(1+\tilde{\gamma}\theta_0
T-\frac{T}{T_1})(\frac{1}{T}+\frac{1}{T_1}){\rm{exp}}(\frac{t-t_0}{T})
-(1-\tilde{\gamma}\theta_0
T+\frac{T}{T_1})(\frac{1}{T}-\frac{1}{T_1})}{(1+\tilde{\gamma}\theta_0
T-\frac{T}{T_1}){\rm{exp}}(\frac{t-t_0}{T})+(1-\tilde{\gamma}\theta_0
T+\frac{T}{T_1})}\right]^2.\label{eq:rho}
\end{equation}
The above solution is valid when $\tilde{\gamma}\neq 0$. For
$\tilde{\gamma}=0$, we need to take the limit case. When
$\tilde{\gamma}\to 0$, the limit of the solution $a(t)$ is got as
\begin{equation}
a(t)=a_0{\rm{exp}}\left[\left(\frac{1}{3}\theta_0T_1+\frac{T_1^2}{T_2^2}
\right)\left(e^{(t-t_0)/T_1}-1\right)-\frac{T_1(t-t_0)}{T_2^2}\right].\label{eq:g0a}
\end{equation}
And we obtain directly
\begin{equation}
\rho(t)=\frac{3}{\kappa^2}\left[\frac{1}{3}\theta_0e^{(t-t_0)/T_1}
+\frac{T_1}{T_2^2}\left(e^{(t-t_0)/T_1}-1\right)\right].\label{eq:g0rho}
\end{equation}
Note that the solution $a(t)$ for $\tilde{\gamma}=0$ has not
possessed the future singularity, the so called Big Rip, in this
case.

\section{Sound speed and EOS parameter}
According to Ref.~\cite{rie04}, the observational consequences are
summarized as follows:
\begin{itemize}
\item They provide the first conclusive evidence
for cosmic deceleration that preceded the current epoch of cosmic
acceleration. Using a simple model of the expansion history, the
transition between the two epochs is constrained to be at $z=0.46
\pm 0.13$.
\item For a flat
universe with a cosmological constant, they measure $\Omega_M=0.29
\pm ^{0.05}_{0.03}$ (equivalently, $\Omega_\Lambda=0.71$).
\item When combined with
external flat-universe constraints including the cosmic microwave
background and large-scale structure, they find $w=-1.02 \pm
^{0.13}_{0.19}$ (and $w<-0.76$ at the 95\% confidence level) for an
assumed static equation of state of dark energy, $p = w\rho$.
\item The constraints are consistent with
the static nature of and value of $w$ expected for a cosmological
constant (i.e., $w_0 = -1.0$, $dw/dz = 0$), and are inconsistent
with very rapid evolution of dark energy.
\end{itemize}

On the basis of the above observational consequences, we suggest
that $\tilde{\gamma}\sim 0$ and the parameter $T_1$ be negative,
if the dark fluid describes the unification of dark matter and
dark energy. Because when $\tilde{\gamma}=0$ and $T_1<0$, we can
obtain
\begin{itemize}
\item The universe can accelerate after the epoch of deceleration.
\item The density approaches to a constant in the late times,
which corresponds to the de Sitter universe, so there is no future
singularity. \item The EOS parameter $w$ approaches to $-1$ when
the cosmic time $t$ is sufficiently large. \item The sound speed
is a real number (see Ref.~\cite{han05} for constraints of sound
speed).
\end{itemize}
Because of these features, we construct a model of dark fluid which
can be seen as a unification of dark energy and dark matter. Using
Eq.~(\ref{eq:f1}), we obtain the relation between $p$ and $\rho$
\begin{equation}
p=(\gamma-1)\rho+p_0+\frac{\kappa}{\sqrt{3}}w_H\sqrt{\rho}
+\frac{\kappa^2}{3}w_{H2}\rho^2-\frac{\kappa^2}{2}w_{dH}(p+\rho),
\end{equation}
that is
\begin{equation}
(1+\frac{\kappa^2}{2}w_{dH})p=(\gamma-1+\frac{\kappa^2}{3}w_{H2}
-\frac{\kappa^2}{2}w_{dH})\rho+\frac{\kappa}{\sqrt{3}}w_H\sqrt{\rho}+p_0.
\end{equation}
\end{widetext}
So the EOS between $p$ and $\rho$ is
\begin{equation}
p=(\tilde{\gamma}-1)\rho-\frac{2}{\sqrt{3}\kappa T_1}
\sqrt{\rho}-\frac{2}{\kappa^2 T_2^2},
\end{equation}
where $\tilde{\gamma}$ is defined as before. Figs. \ref{gamma1}
and \ref{gamma2} show that the parameters $\gamma$, $w_{H2}$, and
$w_{dH}$ can drive $\tilde{\gamma}$ crossing $-1$. In the present
paper, we set $\kappa=1$ and $\theta_0=1$ for simplicity to all
the figures in this paper. The values of other parameters are
given in the legend and the caption of each figure. We especially
consider the following choice of the parameters: $\gamma=0$,
$T_1=-25$, and $T_2=100$.

\begin{figure}[]
\includegraphics{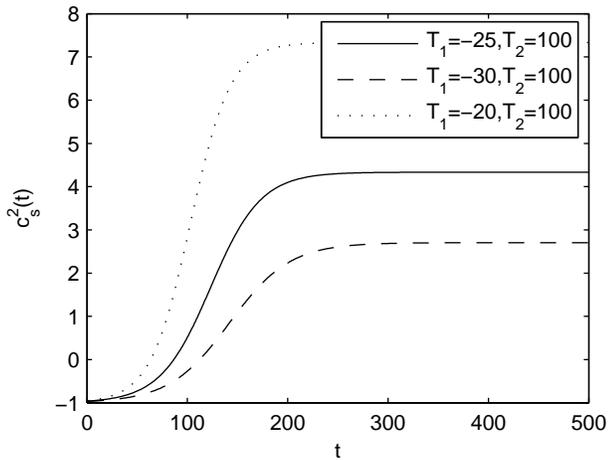}
\caption{\label{g0cs2} The relation between the square of the sound
speed $c_s^2$ and the cosmic time $t$.}
\end{figure}

The square of the sound speed is
\begin{equation}
c_s^2=\frac{\partial
p}{\partial\rho}=\tilde{\gamma}-1-\frac{1}{\sqrt{3}\kappa T_1}
\frac{1}{\sqrt{\rho}}.
\end{equation}
When $\tilde{\gamma}=0$, $T_1$ should be negative if the sound speed
is a real number. The graph of $c_s^2$-$t$ relations is shown in
Fig.~\ref{g0cs2}. We can see that the sound speed approaches to a
constant in the late times.

\begin{figure}[]
\includegraphics{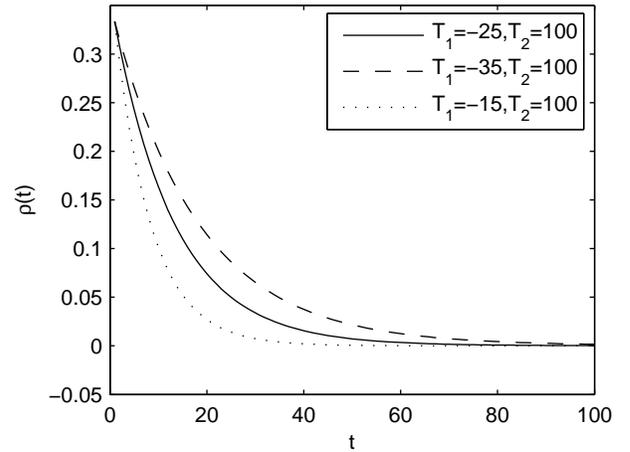}
\caption{\label{g0rho} The relation between the density $\rho$ and
the cosmic time. Note that the density approaches to a constant,
which is not zero.}
\end{figure}

The EOS parameter is
\begin{equation}
w=\frac{p}{\rho}=\tilde{\gamma}-1-\frac{2}{\sqrt{3}\kappa T_1}
\frac{1}{\sqrt{\rho}}-\frac{2}{\kappa^2 T_2^2}\frac{1}{\rho}
\end{equation}
\begin{figure}[]
\includegraphics{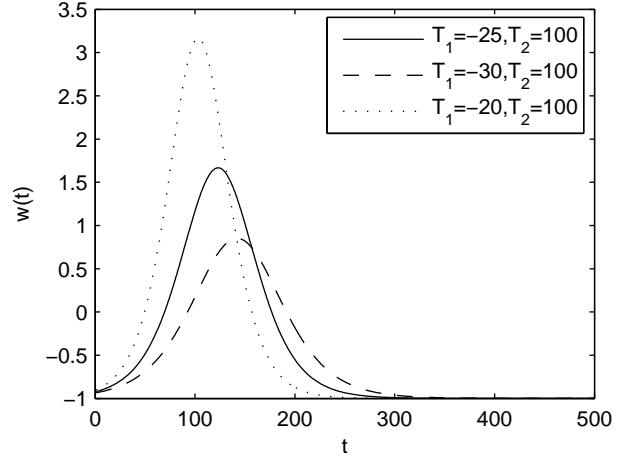}
\caption{\label{g0w} The relation between the EOS parameter
$w=p/\rho$ and the the cosmic time $t$.}
\end{figure}
Fig.~\ref{g0w} shows the $w$-$t$ relation. This figure shows that
$w$ also approaches to a constant in the late times. This is
because the density $\rho$ approaches to a constant after some
time. Because we have already chosen $\tilde{\gamma}=0$, there is
no $w=-1$ crossing. However, if $\tilde{\gamma}$ is around zero,
the crossing may easily occur. Since we chose the parameter $T_1$
to be negative, from Eq.~(\ref{eq:rho}), we can see that the
density approches to a constant
\begin{equation}
\rho=-\frac{3T_1}{\kappa^2 T_2^2}
\end{equation}
after a sufficiently large time, as in Fig.~\ref{g0rho}.

\begin{figure}[]
\includegraphics{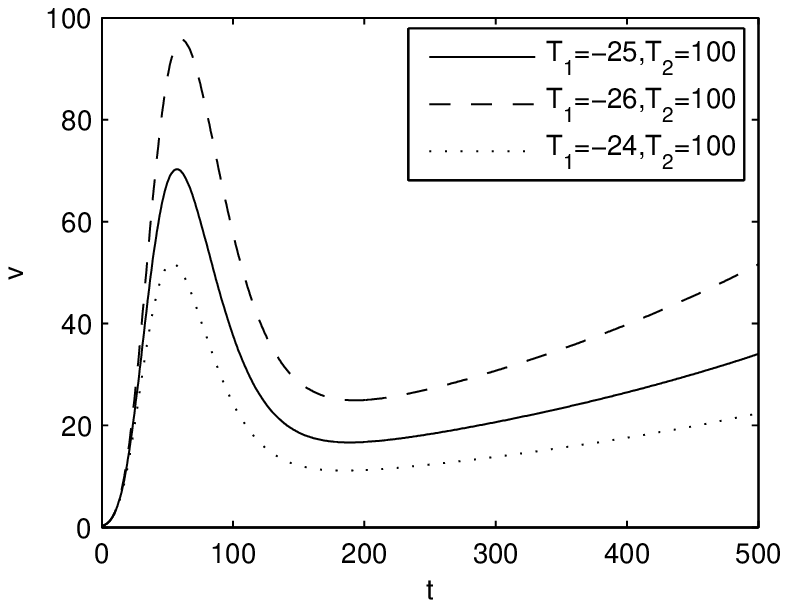}
\caption{\label{g0dota} The relation between the expansion velocity
$v=\dot{a}$ and the cosmic time $t$.}
\end{figure}

\begin{figure}[]
\includegraphics{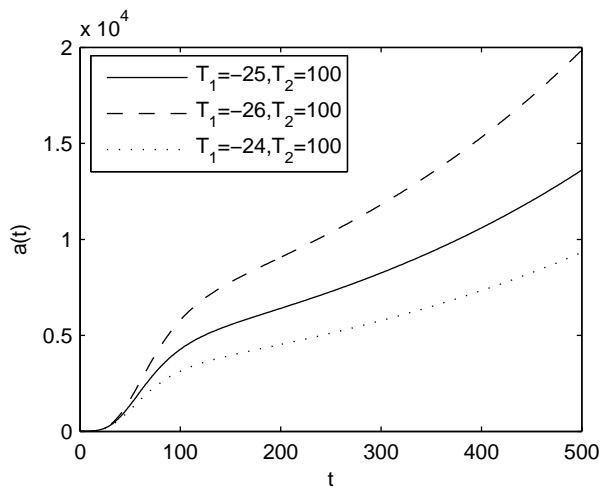}
\caption{\label{g0a} The relation between the scale factor $a$ and
the cosmic time $t$.}
\end{figure}

In order to explain the observations, there should be at least one
term in the right hand side of Eq.~(\ref{eq:main}) causing the
cosmic expansion to accelerate and one forcing the expansion to
decelerate. If we assume that the universe approaches to the de
Sitter space-time in the late times, $(+,-,+)$, $(-,+,+)$, and
$(-,-,+)$ are possible combinations of the signs of the three
terms. It is interesting that Eq.~(\ref{eq:main}) with the signs
$(+,-,+)$ in the right hand side may unify the early-time
inflation, the middle-time deceleration and the late-time
acceleration, which is discussed in the next section.
Fig.~\ref{g0dota} shows that the universe accelerates after an
epoch of deceleration, and Fig.~\ref{g0a} shows the corresponding
evolution of the scale factor.
The case for possible future singularity is considered in our
previous paper \cite{men05}. In Ref.~\cite{noj04b,noj04c,eli04},
they demonstrate that the quantum effects play the dominant role
near/before a big rip, driving the universe out of a future
singularity (or at least, moderating it). It is also interesting
to study the entropy and dissipation \cite{bre04,pri00,her01},
since this EOS may be interpreted as the time-dependent viscosity
case.

\begin{figure}[]
\includegraphics{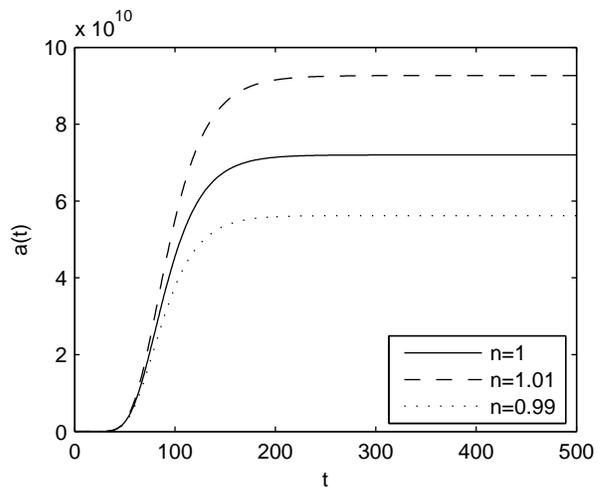}
\caption{\label{g0an} The evolution of the scale factor when
$\gamma=-1$ and $t_c=-25$.}
\end{figure}

The more general EOS, such as the form
\begin{equation}
p_X=-\rho_X-A\rho_X^\alpha-BH^{2\beta},\label{eq:px}
\end{equation}
in Ref.~\cite{cap05}, give more general dynamical equations, which
can be written as
\begin{equation}
\frac{\ddot{a}}{a}=-\frac{3\tilde{\gamma}-2}{2}\frac{\dot{a}^2}{a^2}
+\lambda\left(\frac{\dot{a}}{a}\right)^m
+\mu\left(\frac{\dot{a}}{a}\right)^n+\nu.
\end{equation}
The corresponding coefficients to Eq.~(\ref{eq:px}) are
$\tilde{\gamma}=0$, $\lambda=A(\kappa^2/2)(3/\kappa^2)^\alpha$,
$m=2\alpha$, $\mu=(\kappa^2/2)B$, $n=2\beta$, and $\nu=0$. Now we
only consider a simpler case to illustrate the scale factor
evolution behaviors,
\begin{equation}
\frac{\ddot{a}}{a}=-\frac{3\tilde{\gamma}-2}{2}\frac{\dot{a}^2}{a^2}
+\frac{1}{t_c}\left(\frac{\dot{a}}{a}\right)^n.
\end{equation}
where $t_c$ is a parameter. Fig.~\ref{g0an} shows the evolution of
the scale factor with different $n$.

\section{Interpretations of the model}
\subsection{Unified dark energy}
Since we do not know the nature of either dark energy or dark
matter, maybe they can be regarded as two aspects of a single
fluid. Based on the analysis in the above section, the EOS in our
model can be looked as that of the unified dark energy, since the
universe expansion can accelerate after the epoch of deceleration. The model
in the present paper can also be regarded as the $\Lambda$CDM
model with an additional term, or the $\Lambda$CDM model with bulk
viscosity. The parameter space is enriched in this model. The
$\Lambda$CDM model describes two mixed fluids, and their EOSs are
$p=0$ for dark matter and $p=-\rho$ for dark energy (cosmological constant).
In our model,
the case $\tilde{\gamma}=1$ and $T_1\to\infty$ corresponds to the
$\Lambda$CDM model. However, in the above section, we study a
special choice of the parameters, $\tilde{\gamma}=0$, $T_1=-25$,
and $T_2=100$, which is totally different from the $\Lambda$CDM
model.

The qualitative analysis of Eq.~(\ref{eq:main}) can be easily
obtained if we assume that $H$ is always decreasing during the
cosmic evolution. The three terms in the right hand side of
Eq.~(\ref{eq:main}) are proportional to $H^2$, $H^1$, and $H^0$,
respectively. If we assume $a\propto e^{2t}$, then $H\propto e^t$,
so the proportions of the three terms are $e^{2t}:e^t:1$; if we
assume $a\propto t^{2/3}$, then $H\propto 1/t$, so the proportions
of the three terms are $t^{-2}:t^{-1}:1$. In the early times, the
first term is dominant, which may lead to inflation if
$\tilde{\gamma}\sim 0$. In the roughly middle times, the second term is
dominant, which leads to deceleration if $T_1<0$. In the late times as current,
the third term is dominant, which leads to acceleration like the de
Sitter universe if $T_2$ is a real number. We can also see the
evolution of $\dot{a}(t)$ in Fig.~\ref{g0dota} and $a(t)$ in
Fig.~\ref{g0a}.

In Eq.~(\ref{eq:main}), the term
$\frac{1}{T_1}\frac{\dot{a}}{a}$ describes the effective
viscosity. Since we do not know much about the nature of dark energy and the
bulk viscosity in the universe, so the bulk viscosity can be
regarded as effective, or a contribution as friction term.
 In order to separately study the effect of the three
terms in the right hand side of Eq.~(\ref{eq:main}), if the first,
and the second term are dominant, respectively we have the evolution relations
\begin{subequations}
\begin{eqnarray}
\frac{\ddot{a}}{a}=-\frac{3\gamma-2}{2}\frac{\dot{a}^2}{a^2}
&\Rightarrow& a\frac{dH_\gamma}{da}=-\frac{3\gamma}{2}H_\gamma,\\
\frac{\ddot{a}}{a}=\frac{1}{T_1}\frac{\dot{a}}{a} &\Rightarrow&
a\frac{dH_v}{da}=-H+\frac{1}{T_1}.
\end{eqnarray}
\end{subequations}
\begin{subequations}
The solutions are correspondingly different
\begin{eqnarray}
H_\gamma(z) &=& H_0^2(1+z)^{3\gamma},\\
H_v(z) &=& \left(H_0-\frac{1}{T_1}\right)(z+1)+\frac{1}{T_1}.
\end{eqnarray}
\end{subequations}

\subsection{Mixture of dark energy and dark matter}
Another interpretation is that the EOS describes the dark energy,
which is mixed with the dark matter in the universe media. So we
should concern on the mixture of the dark energy and dark matter, which
requires fine-tuning of the parameters. In the $\Lambda$CDM model of
cosmology, we have
\begin{equation}
\left(\frac{\dot{a}}{a}\right)^2=H_0^2[\Omega_m(1+z)^3+\Omega_\Lambda].
\end{equation}
where $H_0$ is the current value of the Hubble parameter,
$z=a_0/a-1$ is the redshift, $\Omega_m$ and $\Omega_\Lambda$ are the
cosmological density parameters of matter and $\Lambda$-term,
respectively. In our case,
\begin{equation}
\left(\frac{\dot{a}}{a}\right)^2=H_0^2\Omega_m(1+z)^3+(1-\Omega_m)H_d(z)^2,
\end{equation}
where $H_d(z)$ is the solution of the equation
\begin{equation}
aH\frac{dH}{da}=-\frac{3\tilde{\gamma}}{2}H^2+\frac{1}{T_1}H+\frac{1}{T_2^2}.
\label{eq:Ha}
\end{equation}
The solution of the above equation with the initial condition
$H(a_0)=H_0$ is
\begin{equation}
\left|\frac{(H-\frac{1}{3\tilde{\gamma}T_1})^2-\frac{1}{9\tilde{\gamma}^2
T_1^2}-\frac{2}{3\tilde{\gamma}^2
T_2^2}}{(H_0-\frac{1}{3\tilde{\gamma}T_1})^2-\frac{1}{9\tilde{\gamma}^2
T_1^2}-\frac{2}{3\tilde{\gamma}^2 T_2^2}}\right|=(1+z)^{3\gamma}.
\end{equation}
Here we consider a simpler case, $T_2\to\infty$, then
\begin{equation}
a\frac{dH}{da}=-\frac{3\tilde{\gamma}}{2}H+\frac{1}{T_1}.\label{eq:vis}
\end{equation}
The solution is
\begin{equation}
H(a)=\left(H_0-\frac{2}{3\tilde{\gamma}T_1}\right)\left(\frac{a}{a_0}
\right)^{-3\gamma/2}+\frac{2}{3\tilde{\gamma}T_1},
\end{equation}
so $H(z)$ for the dark energy is
\begin{equation}
H_d(z)=\left(H_0-\frac{2}{3\tilde{\gamma}T_1}\right)(z+1)^{3\gamma/2}
+\frac{2}{3\tilde{\gamma}T_1}.
\end{equation}
It is interesting that the above relation can be rewritten as
\begin{equation}
H_d(z)=H_0[\tilde{\Omega}(1+z)^{3\tilde{\gamma}/2}+(1-\tilde{\Omega})],\label{eq:hd}
\end{equation}
where
\begin{equation}
\tilde{\Omega}=1-\frac{2}{3\tilde{\gamma}T_1 H_0}.
\end{equation}
Note that Eq.~(\ref{eq:hd}) is valid if $\tilde{\gamma}\neq 0$, and
for the case $\tilde{\gamma}=0$, directly solving Eq.~(\ref{eq:vis})
gives
\begin{equation}
H_d(z)=H_0\left[1-\frac{1}{T_1 H_0}{\rm{ln}}(z+1)\right].
\end{equation}

We assume the universe media contains two fluids. One is described
by Eq.~(\ref{eq:main}) with $T_2\to\infty$, and another is described
by the simplest pure cosmological constant $\Lambda$. The former may be
regarded as the dark matter with effective viscosity. The $H$-$z$
relation is thus
\begin{equation}
H^2=H_0^2\{\Omega_m[\tilde{\Omega}(1+z)^{3\tilde{\gamma}/2}
+(1-\tilde{\Omega})]^2+(1-\Omega_m)\},
\end{equation}
which can be regarded as the generalized relation of mixed dark
energy and dard matter. We emphasize that solving Eq.~(\ref{eq:Ha})
with $T_2\to\infty$ and writting $H^2=\Omega_m H_d^2+(1-\Omega_m)H_0^2$
is not equivalent to directly solving Eq.~(\ref{eq:Ha}) (see
Appendix for details).

If the universe media can be perceived as the mixture of matter,
radiation, $\Lambda$ term, and with effective viscosity, by ignoring curvature
contribution as favored from WMAP data. The total density is
\begin{equation}
\rho=\rho_m+\rho_r+\rho_\Lambda+\rho_v,
\end{equation}
where the subscripts denote the matter, radiation, $\Lambda$, and
effective viscosity components, respectively. Since $\rho\propto H^2$, we have
\begin{equation}
H(z)^2=\Omega_m H_m^2+\Omega_r H_r^2+\Omega_\Lambda
H_\Lambda^2+\Omega_v H_v^2.
\end{equation}

\subsection{Effective viscosity model}
The third interpretation is a new model called effective viscosity
model, which may be a most significant result in this paper. We
assume the universe media can be described by only a single fluid, which corresponds
to the matter described by the EOS of $p=0$, and with an effective
constant viscosity. In this model, it is the effective viscosity
that causes the cosmic expansion acceleration without by introducing a cosmological constant,
which is totally different from the $\Lambda$CDM model. We rewrite
Eq.~(\ref{eq:hd}) with $\tilde{\gamma}=1$ as
\begin{equation}
H(z)=H_0[\Omega_m(1+z)^{3/2}+(1-\Omega_m)]
\end{equation}
There is one adjustable parameter in this model. In the next
section we will show that this model can fit the SNe Ia data at an
acceptable level.

\section{Data fitting of the effective viscosity model}
\begin{figure}[]
\includegraphics{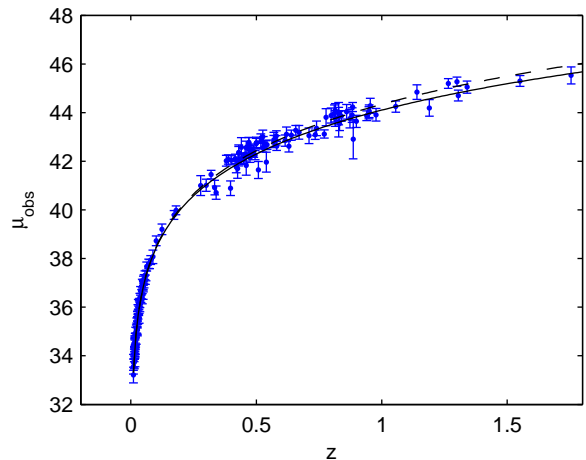}
\caption{\label{thobs} The dependence of luminosity on redshift
computed from the effective viscosity model. The solid and dashed
lines correspond to $\Omega_m=0.3$ and $\Omega_m=0.5$, respectively.
The dots are the observed data.}
\end{figure}

The observations of the SNe Ia have provided the first direct evidence of the
accelerating expansion for our current universe. Any model attempting to explain the
acceleration mechanism should be consistent with the SNe Ia data implying results,
as a basic requirement. The
method of the data fitting is illustrated in Ref.~\cite{ben05}. The
observations of supernovae measure essentially the apparent
magnitude $m$, which is related to the luminosity distance $d_L$ by
\begin{equation}
m(z)={\cal M}+5{\rm{log}}_{10} D_L(z),
\end{equation}
where $D_L(z)\equiv(H_0/c)d_L(z)$ is the dimensionless luminosity
distance and
\begin{equation}
d_L(z)=(1+z)d_M(z),
\end{equation}
where $d_M(z)$ is the comoving distance given by
\begin{equation}
d_M(z)=c\int_0^z\frac{1}{H(z')}dz'.
\end{equation}
Also,
\begin{equation}
{\cal
M}=M+5{\rm{log}}_{10}\left(\frac{c/H_0}{1{\rm{Mpc}}}\right)+25,
\end{equation}
where $M$ is the absolute magnitude which is believed to be constant
for all supernovae of type Ia. We use the 157 golden sample of
supernovae data compiled by Riess \textit{et al.} \cite{rie04} to
fit our model. The data points in these samples are given in terms
of the distance modulus
\begin{equation}
\mu_{obs}(z)\equiv m(z)-M_{obs}(z).
\end{equation}
The $\chi^2$ is calculated from
\begin{equation}
\chi^2=\sum_{i=1}^{n}\left[\frac{\mu_{obs}(z_i)-{\cal M'}
-5{\rm{log}}_{10}D_{Lth}(z_i;c_\alpha)}{\sigma_{obs}(z_i)}\right]^2.
\end{equation}
where ${\cal M'}={\cal M}-M_{obs}$ is a free parameter and
$D_{Lth}(z_i;c_\alpha)$ is the theoretical prediction for the
dimensionless luminosity distance of a supernovae at a particular
distance, for a given model with parameters $c_\alpha$. We
consider the generalized $\Lambda$CDM model as referred to in the previous
section and perform a best-fit analysis with the minimization of the
$\chi^2$, with respect to ${\cal M'}$, $\Omega_m$. Fig.~\ref{thobs}
shows that the theoretical curve fits the observational data at an
acceptable level, with only one adjustable parameter $\Omega_m$
except the today's Hubble parameter $H_0$.

\section{Discussion and conclusion}
We investigate a parameterized effective EOS of dark fluid in the
cosmological evolution. With this general EOS, the dynamical
equation of the scale factor is completely integrable and an exact
solution for Einstein's gravitational equation with FRW metric is obtained. The
parameters $\gamma$, $p_0$, $w_H$, $w_{H2}$, and $w_{dH}$ can be
reduced to three condensed parameters $\tilde{\gamma}$, $T_1$, and $T_2$.
Three interpretations to this model are proposed in this paper:
\begin{itemize}
\item This EOS can be regarded
as a unification of the dark energy and dark matter, so there is a
single fluid to show functions in the universe. In this case, we prefer to the choice of
the parameters: $\tilde{\gamma}\sim 0$, $T_1<0$, and $T_2^2>0$.
\item This
EOS describes the dark energy, which is mixed with the dark matter
in the universe media; or this EOS describes the dark matter with
viscosity, which is mixed with the dark energy from the
$\Lambda$-term.
\item The universe media contains a single fluid, which corresponds
to the matter described by the EOS of $p=0$, with an effectively
constant viscosity. It is the effective viscosity that causes the
cosmic expansion acceleration without by introducing a cosmological constant. In this case,
we prefer to the choice of the parameters: $\tilde{\gamma}\sim 1$,
$T_1>0$, and $T_2=0$.
\end{itemize}
Different choices of the parameters may lead to several fates of the
cosmological evolution. We especially study the choices for the
parameters of $\tilde{\gamma}=0$ and $T_1<0$ and the unified dark
energy in the first interpretation case. We presents a generalized relation
of $H$-$z$ compared with the $\Lambda$CDM model. We show that the
matter described by the EOS of $p=0$ plus with effective viscosity and
without introducing the cosmological constant can fit the observational data
well, so the effective viscosity model may be an alternative
candidate to explain the late-time accelerating expansion universe.

\appendix
\section{Remarks on the $\Lambda$-term invovled}
Directly solving Eq.~(\ref{eq:Ha}) with the EOS $p=(\gamma-1)\rho$
gives
\begin{equation}
H(z)=\left(H_0^2-\frac{2}{3\tilde{\gamma}^2
T_2^2}\right)(1+z)^{3\gamma}+\frac{2}{3\tilde{\gamma}^2 T_2^2},
\end{equation}
which can be rewritten as
\begin{equation}
H^2=H_0^2[\Omega_m(1+z)^{3\gamma}+(1-\Omega_m)],\label{eq:w}
\end{equation}
where $\Omega_m=1-\frac{2}{3\tilde{\gamma}^2 T_2^2 H_0^2}$. Solving
the Friedmann equations with the EOS $p=(\gamma-1)\rho$ without the
$\Lambda$-term gives $H_x^2=H_0^2(1+z)^{3\gamma}$. On the other hand,
concerning on the mixture of the dark energy and dark matter, we can
write $H^2=\Omega_m H_x^2+(1-\Omega_m)H_0^2$, which is exactly the
same as Eq.~(\ref{eq:w}). However, the following two methods are not
equivalent except for some very special cases: (i) Using the EOS
$p=f(\rho)$ to solve the Friedmann equations with the
$\Lambda$-term, we obtain the $H$-$z$ relation. (ii) Using the EOS
$p=f(\rho)$ to solve the Friedmann equations without the
$\Lambda$-term, we obtain $H_x(z)$ and write
\begin{equation}
H^2=\Omega_m H_x(z)^2+(1-\Omega_m)H_0^2.
\end{equation}
Generally the above $H$-$z$ relation is not equivalent to what is
obtained in (i).

\section*{ACKNOWLEDGEMENTS}
X.H.M. is very grateful to Profs. S.D. Odintsov and I. Brevik for lots of
helpful comments with reading the main contents. This work is supported partly by NSF and Doctoral
Foundation of China.

\end{document}